\documentclass[
  aps,
  pra,
  amsmath,amssymb,
  reprint, 
  showpacs
]{revtex4-1}
\usepackage{graphicx,color}
\usepackage{amsmath}
\usepackage{natbib}
\usepackage{epsfig}
\begin{document}

\title{Fingerprints of different interaction mechanisms on the collective modes in complex (dusty) plasmas}

\author{Sergey A. Khrapak,$^{1,2,3}$ Boris A. Klumov,$^{1,3,4}$ and H. M. Thomas$^2$}
\affiliation{$^1$Aix Marseille University, CNRS, Laboratoire PIIM, Marseille, France}
\affiliation{$^2$Forschungsgruppe Komplexe Plasmen, Deutsches Zentrum f\"{u}r Luft- und Raumfahrt,
Oberpfaffenhofen, Germany}
\affiliation{$^3$Joint Institute for High Temperatures, Russian Academy of Sciences, Moscow, Russia}
\affiliation{$^4$L.D. Landau Institute for Theoretical Physics, Russian Academy of Sciences, Moscow, Russia}

\begin{abstract}
In this paper we discuss the relations between the exact shape of interparticle interactions in complex (dusty) plasmas and the dispersion relation of the longitudinal collective mode. Several  representative repulsive potentials, predicted previously theoretically, are chosen and the corresponding dispersion relations are calculated using the quasi-crystalline approximation. Both weakly coupled and strongly coupled regimes are considered. It is shown that the long-wavelength portions of the dispersion relations are sensitive to the long-range asymptote of the interaction potential. This can be used to discriminate between different interaction mechanisms operational in complex plasmas experimentally. Main requirements are briefly discussed.   
\end{abstract}

\date{\today}
\maketitle

\section{Introduction}

Complex (dusty) plasmas consist of weakly ionized gas (conventional plasma) and charged macroscopic (dust) particles~\cite{VladimirovPR2004,FortovUFN2004,FortovPR2005,IshiharaJPD2007,ShuklaRMP2009,CPBook}. 
In laboratory conditions, the (floating) potential of the particle surface is normally set by the condition that the collected electron and ion fluxes balance each other on average. Since electrons are much more mobile than ions, the surface potential is negative and is of the order of the electron temperature (in energy units). This ensures that most of the electrons are reflected from the potential barrier between the particle surface and the surrounding plasma in order for the electron and ion fluxes to be equal. Given that the relation between the charge and the surface potential of a small particle in a plasma is close to that in vacuum, the typical values of particle charge are on the order of $10^3$-$10^4$ elementary charges for particles in the micron-size range and eV-range electron energy~\cite{FortovPR2005,KhrapakPRE2005,KhrapakCPP2009}. Naturally, the highly charged particles interact with each other electrically, and the electrical interaction energy can often be remarkably higher compared to their kinetic energy. This is the main reason why the particle component usually forms condensed liquid and solid phases and exhibits transitions between these phases~\cite{ChuPRL1994,ThomasPRL1994,Hayashi1994,Melzer1994,
ThomasNature1996,MorfillPRL1999,NefedovNJP2003,KlumovPPCF2009,KlumovUFN2010,
KhrapakPRL2011,KhrapakPRE2012}.

Complex plasma can be viewed as a classical system of individually visible strongly interacting particles~\cite{CPBook,BonitzRPP2010}. Relatively weak damping from the plasma background (dominated by the neutral gas) and the absence of hydrodynamic interactions make complex plasmas very suitable models to understand atomic and molecular systems beyond the limits of continuous media. Not surprisingly, it has been recently recognized that this new class of soft matter -- the “plasma state of soft matter” \cite{MorfillRMP2009,ChaudhuriSM2011} -- can be used (complementary to other soft matter systems like colloids, granular medium, etc.) to investigate a broad range of important fundamental processes (equilibrium and non-equilibrium phase transitions, phase separation in multi-component systems, self-organizations, rheology, waves, transport, etc.) at the most fundamental individual particle level. 

As in most other interacting particle systems, the exact shape of the interaction potential between the particles is a key factor determining the rich variety of physical phenomena involved. In complex plasmas interactions are not fixed, but can vary considerably. In particular, the important property of complex plasmas - their thermodynamic openness (associated with continuous exchange of matter and energy between the particles and the surrounding plasma) - results in a remarkable diversity of interaction mechanisms. This diversity is not a problem, but rather an advantage: It widens the range of phenomena accessible for detailed investigation. The problem is the current state of our understanding: While considerable progress has been made in the last decade to understand basic properties of plasma-particle and particle-particle interactions theoretically, there is a significant lack regarding experimental confirmations of these findings. 

The main purpose of this paper is to discuss one of the
possible relations between the exact shape of the interparticle interactions and phenomena relatively easily observable in experiments. In particular, we perform systematic analysis on how deviations from the usually assumed Yukawa (Debye-H\"uckel or screened Coulomb) potential can affect the dispersion relations of collective modes in complex plasmas. In the present paper we limit ourselves to the longitudinal mode in three-dimensional complex plasmas with repulsive interactions between the particles. Generalizations to the two-dimensional situations as well as attractive potentials are relatively straight-forward and may be addressed in future work. The paper is organized as follows: In section~\ref{interactions} we provide a brief overview of the interaction mechanisms, which may operate in complex plasmas, according to the current theoretical understanding. In Section~\ref{QCA_Sec} we introduce the quasi-crystalline approximation used to calculate the dispersion relation of the longitudinal mode associated with the presence of charged particles in a plasma. In Section~\ref{Model_pot} we discuss the model potentials, which can represent the actual interactions in complex plasmas under different conditions. The dispersion relations for these potentials are then calculated and the results are presented in Section~\ref{DR_Sec} for both weakly coupled and strongly coupled regimes. The effect of neutral gas damping is briefly considered in the same section. This is followed by discussion and conclusion in Section~\ref{Disc}.

\section{Brief overview of the interaction mechanisms in complex plasmas}\label{interactions}

The study of interactions between the particles immersed in a plasma is a basic physical problem with many applications ranging from astrophysical topics to technological plasma applications. One naturally cannot avoid dealing with this problem in complex plasmas, since interparticle interactions affect or determine most of the observable phenomena. Considerable progress has been achieved, although in large part from the theoretical perspective, in the last couple of decades to understand particle-particle interactions and their diversity in complex plasmas. Below we briefly summarize the main results obtained so far. The focus is on the  interactions in the three-dimensional (3D) case.    

(i)	The conventional concept of the exponentially screened Coulomb (i.e. Debye-H\"uckel or Yukawa) potential (familiar from conventional plasmas and colloidal suspensions~\cite{BelloniJPCM2000}), where the screening comes from the equilibrium redistribution of plasma electrons and ions in the vicinity of the test charge, can only be used as a very rough zero approximation. The actual interactions between the particle and surrounding plasma involves more than only screening. In particular, continuous absorption (loss) of plasma on the particle surface results in non-equilibrium (non-Boltzmann) character of electron and ion distributions. 

(ii) To be more specific, continuous plasma absorption on the particle surface implies continuous plasma fluxes towards the particles. In the absence of plasma production and loss, conservation of these fluxes results in a power law decay of the electrical potential and similar scaling of the interaction between a pair of particles. In the collisionless situation (ion mean free path is much longer than the plasma screening length) the long-range asymptote of the electrical potential around an individual particle scales as $\phi_{\rm LR}(r)\propto r^{-2}$. This result is well known in the context of spherical Langmuir probes in plasmas~\cite{Laframboise,Alpert} and also in the context of dusty plasmas~\cite{TsytovichUFN,Allen2000,Lampe2000}. In the highly collisional (continuum) limit, the electrical potential decays as $\phi_{\rm LR}(r)\propto r^{-1}$~\cite{Zobnin2000,KhrapakPoP2006,Filippov2008}. In the most interesting for practical applications intermediate regime (moderate collisionality) both scalings are present~\cite{Filippov2008,KhrapakPRL2008,KhrapakPRL2009,ChaudhuriIEEE2010}
and the long-range asymptote of the potential can be presented as $\phi_{\rm LR}\propto c_1/r+c_2/r^2 $, where the parameters $c_1$ and $c_2$ can be in principle adjusted by appropriate variations of plasma density, neutral gas pressure, particle size, etc.  This can potentially be used to ``design'' a required interaction for a particular problem to investigate.

(iii)	Electron and ion production (ionization) and loss (e.g. recombination) in a plasma surrounding particles can result in the emergence of two dominating asymptotes, both having Yukawa form -- the double-Yukawa repulsive potential~\cite{FilippovJETP2007,KhrapakPoP2010}.  The screening length scales can be very different: The first (short-range) term is normally determined by the classical mechanism of Debye-H\"uckel screening, the effective screening length is of the order of the Debye radius. The magnitude of the second (long-range) term is merely controlled by the balance between the plasma production and loss, which typically results in a screening length considerably longer than the Debye radius. Recent studies of fluid-fluid demixing in binary complex plasmas provide a relevant example where the appearance of such two-scale interaction can play a crucial role~\cite{WysockiPRL2010,SutterlinPPCF2010}.  

(iv)	If the particles are not only absorbing electrons and ions from the plasma, but emit electrons (e.g. due to thermionic, photoelectric, or secondary electron emission), their charge can become less negative, and under certain conditions even reach positive values. In this regime a possibility of long-range electrical attraction between positively charged particles has been predicted theoretically~\cite{DelzannoPRL2004,KhrapakPRL2007}. The resulting potential has either a double-Yukawa shape with attractive long-range term~\cite{DelzannoPRL2005}, or Yukawa plus attractive Coulomb long-range asymptote in the highly collisional continuum limit~\cite{KhrapakPRL2007,DyachkovJETP2008}.      

(v)	Besides electrical effects, there exist other mechanisms, associated with complex plasma openness, which can contribute to interparticle interactions. For instance, constant plasma absorption on the particle surfaces gives rise to the so-called "ion shadowing" interaction (sometimes also called ``Lesage gravity'') which basically represents the plasma drag that one particle experiences as a consequence of the plasma flux directed to another neighbouring particle and vice versa~\cite{TsytovichUFN,IgnatovPPR1996}. This attraction mechanism is to some extent analogous to depletion interaction in colloids~\cite{LekkerkerkerBook}, although the detailed physics is different. The ion shadowing interaction exhibits Coulomb-like asymptote ($\propto r^{-1}$) at large interparticle separation~\cite{TsytovichUFN,IgnatovPPR1996,KhrapakPRE2001,KhrapakPoP2008}.    

(vi)	A similar mechanism can be associated with the neutral component, provided the particle surface temperature is different from the temperature of the surrounding neutral gas so that net momentum fluxes between the particle and neutral gas components exist~\cite{TsytovichUFN}. Since the particle surface temperature is determined by a complicated balance of heating and cooling mechanisms such as electron and ion collection and recombination on the surface, exchange of energy with neutrals, plasma and particle radiation, chemical reactions on the surface, it is natural to expect some temperature difference (normally one expects that the surface temperature is somewhat higher than that of the neutral gas)~\cite{Daugherty1993,Swinkels2000,KhrapakTemp}. 

(vii) In addition, exciting possibilities to design new interaction classes tunable to various isotropic/anisotropic and repulsive/attractive forms, by applying external ac fields of various polarizations have been discussed~\cite{IvlevPRL2008,KompaneetsPoP2009,ThomaIEEE2010,MiticPoP2013}.

Thus, the interaction mechanisms in complex plasmas are very diverse, providing an intriguing opportunity to design repulsive and attractive interactions of various required shapes. One of the main obstacles at this point is the absence of reliable direct experimental evidence of the relevance of the mechanisms considered above.
Here we discuss an experimental tool which can be used to fill this gap. In particular, we propose to use the fact that the dispersion of collective modes in the system of interacting particles is rather sensitive to the exact shape of the interaction potential. Using several representative examples, relevant to complex  plasmas, we demonstrate how the dispersion relation of the longitudinal waves reacts to the variations in the interparticle interactions. The quasi-crystalline approximation, also known as the quasi-localized charge approximation, is used for this purpose. This allows us to treat simultaneously both weakly coupled gaseous and strongly coupled fluid regimes (crystalline phase is not considered), which can occur under typical natural and experimental conditions. The obtained results can be used to design dedicated experiments aiming at verifying the existing interaction mechanisms in complex plasmas. As pointed out in the introduction, in this paper we only consider repulsive interactions. We plan to present the results for attractive interactions in a later paper.

\section{Quasi-crystalline approximation}\label{QCA_Sec}

The quasi-crystalline approximation (QCA) was proposed in Ref.~\cite{Hubbard1969} and further detailed in Ref.~\cite{Takeno1971}. This theoretical approach can be regarded as a generalization of the phonon theory of solids or, alternatively, as a generalization of the random phase approximation. In its simplest version, the particles forming liquid are assumed stationary (i.e. like in cold amorphous solid) but the system is characterized by a liquid-like order, measured in terms of the isotropic radial distribution function (RDF) $g(r)$. The linear response of such disordered system can be approximately calculated and related to the frequencies of the collective modes~\cite{Hubbard1969}. Comparable expressions can also be obtained from the analysis of the fourth frequency moment~\cite{deGennes1959}. In the context of plasma physics, similar approach is known as the qusilocalized charge approximation (QLCA)~\cite{GoldenPoP2000}.
In last decades the QLCA approach has been successively applied to describe collective modes in various strongly coupled plasma systems. In particular, this includes one-component-plasma~\cite{GoldenPoP2000,KhrapakOCP2D} and complex plasmas with Yukawa interactions~\cite{RosenbergPRE1997,KalmanPRL2000,OhtaPRL2000,
KalmanPRL2004,DonkoJPCM2008}, in both 3D and 2D situations. Applications to the Lennard-Jones-like and inverse-power-law interactions have also been briefly discussed~\cite{RosenbergCPP2015,KhrapakJCP2016}. 

In the QCA model the dispersion relations are related to the interparticle interaction potential $V(r)$ and the equilibrium radial distribution function $g(r)$ of particles. The compact expression for the longitudinal mode dispersion relation in a single component system is
\begin{equation}\label{QCA}
\omega^2=\frac{n}{m}\int\frac{\partial^2 V(r)}{\partial z^2} g(r) \left[1-\cos(kz)\right]d{\bf r},
\end{equation}
where $\omega$ is the frequency,  $k$ is the wave number, $n$ is the density, $m$ is the particle mass, and $z=r\cos\theta$ is the direction of the propagation of the longitudinal wave.

Below we take several representative examples of repulsive interactions, operational in complex plasmas under different conditions, and calculate the longitudinal dispersion relation with the help of Eq.~(\ref{QCA}). We are then able to identify how the deviations from the simple Yukawa form can affect the dispersion curves and whether this can be potentially used to discriminate between different interactions in experiments.            

\section{Model interaction potentials}\label{Model_pot}

Taking into account the discussion in Section~\ref{interactions}, we have chosen two distinct model interaction potentials for this study. The first is the repulsive double Yukawa potential
\begin{equation}\label{pot1}
V(r)=\frac{Q^2}{r}\left[\epsilon_1\exp(-r/\lambda_1)+\epsilon_2\exp(- r/\lambda_2)\right],
\end{equation}
where $Q$ is the particle charge, $\epsilon_{1,2}$ are positive coefficients ($\epsilon_{1,2}\leq 1$), and $\lambda_{1,2}$ are the effective screening lengths.
This interaction potential has been predicted for the case when electron and ion production (ionization) and loss are significant in a plasma surrounding the particles~\cite{FilippovJETP2007,KhrapakPoP2010}. The functional form (\ref{pot1}) is also advantageous, because it includes single Coulomb ($\lambda_1,~~\lambda_2\rightarrow \infty$, $\epsilon_1+\epsilon_2=1$) and Yukawa ($\epsilon_1=1$, $\epsilon_2=0$, $\lambda_1=\lambda_{\rm D}$) limiting cases. It also describes electrical interactions in highly collisional plasmas (Yukawa plus long range Coulomb asymptote)~\cite{Zobnin2000,KhrapakPoP2006,Filippov2008,BystrenkoPRE2003,KhrapakJAP2007}.
Below we apply the following restriction, $\epsilon_1+\epsilon_2=1$ in order to recover the Coulomb short-range asymptote near the particle origin (particles are treated as point-like throughout the paper).  

The parameters $\epsilon_{1,2}$ and $\lambda_{1,2}$ can in principle vary in a relatively wide range, depending on exact mechanisms responsible for the appearance of the second term in Eq.~(\ref{pot1}) as well as other plasma parameters. We adopt the three following parameter sets for this study. {\it Case 1}: $\epsilon_1=\epsilon_2=0.5$, $\lambda_1=0.7\lambda_{\rm D}$, $\lambda_2=6.3\lambda_{\rm D}$. This choice corresponds to an exemplary calculation of a test charge shielding taking into account plasma production and loss processes~\cite{KhrapakPoP2010}. In particular, these numbers were obtained for  isothermal plasma with ambipolar losses dominating over the losses due to the three-body recombination for a reduced ionization rate equal to unity (see Fig. 1 from Ref.~\cite{KhrapakPoP2010} for details).
{\it Case 2}:  $\epsilon_1=0.8$, $\epsilon_2=0.2$, $\lambda_1=\lambda_{\rm D}$, $\lambda_2=10\lambda_{\rm D}$. This parameter set is close to that used to model the kinetics of fluid-fluid demixing in binary complex plasmas, observed experimentally using PK-3 Plus laboratory on board the International Space Station~\cite{WysockiPRL2010}. 
{\it Case 3}: $\epsilon_1=0.5$, $\epsilon_2=0.5$, $\lambda_1=\lambda_{\rm D}$, $\lambda_2=\infty$. This shape corresponds to the Yukawa potential with the unscreened Coulomb long-range asymptote. Such situation is relevant to either electrical interactions in a highly collisional plasma~\cite{Zobnin2000,KhrapakPoP2006,Filippov2008,BystrenkoPRE2003,KhrapakJAP2007}, or to a plasma with developed ionization, when all losses are associated with the ambipolar diffusion~\cite{KhrapakPoP2010}. The parameters adopted here are representative for electrical interactions in highly collisional isothermal plasma~\cite{KhrapakPoP2006,KhrapakJAP2007}.


\begin{figure}
\includegraphics[width=8.5cm]{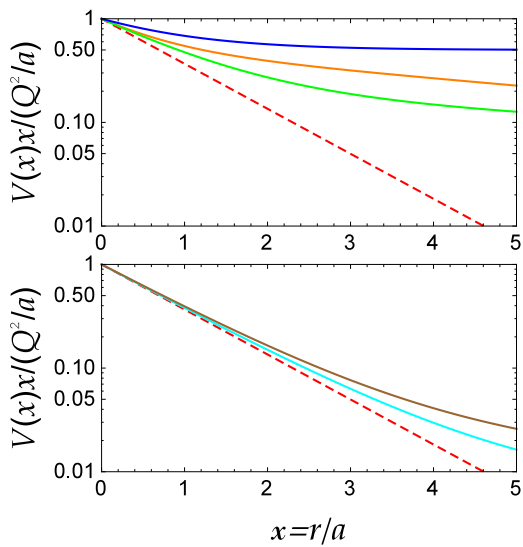}
\caption{(color online) Reduced model potentials used in this study. The top panel shows the double Yukawa repulsive potentials corresponding to the Case 1 (orange), Case 2 (green), and Case 3 (blue). The bottom panel shows the Yukawa potential with long-range inverse second power asymptote, corresponding to the Case 4 (cyan) and Case 5 (olive). The dotted red line in both figures shows the conventional single Yukawa potential (\ref{Yukawa}). }
\label{Fig1}
\end{figure}

The second model potential we investigate here mimics the interaction between two collecting particles in collisionless plasmas,
\begin{equation}\label{pot2}
V(r)=\frac{Q^2}{r}\left[(1-\epsilon)e^{-r/\lambda_D}+(\epsilon\lambda_{\rm D}/r)\left(1-e^{-r/\lambda_{\rm D}}\right)\right],
\end{equation}   
where the screening is described by conventional Debye-H\"uckel scenario with the screening length $\lambda_{\rm D}$ and the (repulsive) long-range asymptote of the potential decays as $\propto r^{-2}$. The model form chosen ensures $V(r)\simeq Q^2/r$
at short separations between the particles and $V(r)\simeq \epsilon \lambda_{\rm D}Q^2/r^2$ in the limit of large separation. The actual magnitude of the long-range asymptote can be estimated~\cite{TsytovichUFN,FortovPR2005,KhrapakPRL2008} as $U_{\rm LR}\simeq Q^2a/2r^2$, which immediately yields $\epsilon=a/2\lambda_{\rm D}$. In the majority of experiments the particle radius is sufficiently small, $a\ll \lambda_{\rm D}$. Therefore, here we take the following two representative values, {\it Case 4}: $\epsilon = 0.05$; and {\it Case 5}: $\epsilon = 0.1$.

In the following, the normalized units for the distance are used, $x=r/a$, where  $a=(4\pi n/3)^{-1/3}$ is the characteristic interparticle distance. In addition, we         
set the screening parameter $\kappa=a/\lambda_{\rm D}$ to unity ($\kappa =1$) for all the cases considered. For convenience, the interaction types and the corresponding sets of parameters are summarized in Table~\ref{TabI}.  

\begin{table}
\caption{\label{TabI} Summary of the model interaction potentials considered in this study (Cases 1 - 5).}
\begin{ruledtabular}
\begin{tabular}{lll}
Case & Functional form & Parameters   \\ \hline
1 & Eq.~(\ref{pot1}) &  $\epsilon_1=\epsilon_2=0.5$, $\lambda_1=0.7\lambda_{\rm D}$, $\lambda_2=6.3\lambda_{\rm D}$     \\
2 & Eq.~(\ref{pot1}) &   $\epsilon_1=0.8$, $\epsilon_2=0.2$, $\lambda_1=\lambda_{\rm D}$, $\lambda_2=10\lambda_{\rm D}$     \\
3 & Eq.~(\ref{pot1}) &   $\epsilon_1=\epsilon_2=0.5$, $\lambda_1=\lambda_{\rm D}$, $\lambda_2=\infty$    \\
4 & Eq.~(\ref{pot2}) &   $\epsilon=0.05$   \\
5 & Eq.~(\ref{pot2}) &   $\epsilon=0.1$   \\
\end{tabular}
\end{ruledtabular}
\end{table} 

The chosen model potentials are plotted in Fig.~\ref{Fig1}, where they are also compared with the conventional single Yukawa potential. Of course, the chosen examples do not cover all the possibilities of interactions between the particles in complex plasmas. In particular, we remind that in this paper we consider only repulsive interactions. Nevertheless, the examples chosen are representative enough to make some conclusions about how the deviations from the conventional single-Yukawa form can affect the dispersion of the longitudinal waves.   
    
\section{Dispersion relations} \label{DR_Sec}   

\subsection{Weakly coupled regime}

The QCA theory was originally developed as a tool to describe collective motion in liquids. However, it was also pointed out that in the special case of a cold crystalline solid it yields the conventional phonon-dispersion relation. In the opposite limit, when correlations between the particle positions can be completely neglected, the QCA reduces to the usual random phase approximation theory of plasmas~\cite{Hubbard1969}. Thus, the region of the applicability of the QCA is wider than seems appropriate at first. Here we first apply QCA to describe dispersion relations of complex plasmas at weak coupling. It is appropriate to start by analysing the corresponding dispersion relation for a single-Yukawa potential 
\begin{equation}\label{Yukawa}
V(r)=\frac{Q^2}{r}\exp(-r/\lambda_D),
\end{equation} 
assuming weak correlations (weak coupling) between the particles. We substitute the radial distribution function $g(r)=1$ into Eq.~(\ref{QCA}) along with the potential (\ref{Yukawa}) to get (for details of the calculation see Appendix)
\begin{equation}\label{DAW}
\omega^2=\frac{\omega_{\rm p}^2q^2}{q^2+\kappa^2},
\end{equation}
where $\omega_{\rm p}=\sqrt{4\pi Q^2n/m}$ is the plasma frequency associated with the charged particle component and $q=ka$ is the reduced wave number. The dispersion relation of this mode, known as the dust-acoustic-wave (DAW), was originally derived using the conventional fluid approach for a multi-component plasma in Ref.~\cite{Rao1990}. Note, that in the limit of infinite screening length, $\kappa\rightarrow 0$, we recover the conventional plasmon dispersion of the classical 3D one-component-plasma (or, equivalently, the Langmuir wave),
\begin{equation}
\omega\simeq \omega_{\rm p}.
\end{equation}  
The dispersion relation (\ref{DAW}) exhibits the following properties: In the long-wavelength limit ($q\lesssim 1$) dispersion is acoustic-like ($\omega\propto q$) with the acoustic velocity
\begin{equation}
c_{\rm DAW}= \omega_{\rm p}\lambda_{\rm D},
\end{equation} 
usually referred to as the dust-acoustic velocity. At shorter wavelengths, the frequency increases monotonically, approaching the short-wavelength asymptote $\omega\simeq \omega_{\rm p}$. 

The generalization to the double-Yukawa potential is trivial. Using the additivity property of the QCA in the weak coupling limit we immediately get for the potential~(\ref{pot1})
\begin{equation}\label{Disp_DY}
\omega^2=\epsilon_1\frac{\omega_{\rm p}^2q^2}{q^2+\kappa_1^2}+\epsilon_2\frac{\omega_{\rm p}^2q^2}{q^2+\kappa_2^2},
\end{equation} 
where $\kappa_{1,2}=a/\lambda_{1,2}$. 
Comparable expressions for the dispersion relation in a weakly coupled complex plasma with double-Yukawa interactions between the particles were previously obtained using the method of moments and the hydrodynamic approach in Refs.~\cite{FilippovJETPLett2010,FilippovPLA2011}. We see that QCA provides a particularly simple route to derive this dispersion.

In the short-wavelength limit, the dispersion relation 
(\ref{Disp_DY}) behaves similarly to the single Yukawa case, $\omega\simeq\omega_{\rm p}$ (we remind that $\epsilon_1+\epsilon_2=1$), which stems from the short range Coulombic asymptote of the interaction potential. In the long-wavelength limit we recover the acoustic branch if both $\kappa_1$ and $\kappa_2$ are non-zero. The acoustic velocity is
\begin{equation}
c_{\rm s}=\omega_{\rm p}\sqrt{\epsilon_1\lambda_1^2+\epsilon_2\lambda_2^2}.
\end{equation}
Since normally $\lambda_1\simeq \lambda_{\rm D}$ and $\lambda_2\gg\lambda_{\rm D}$, this acoustic velocity can significantly exceed the conventional $c_{\rm DAW}$. If $\lambda_{2}=\infty$ (and $\kappa_2=0$), as in the Case 3, the long-wavelength behaviour is non-acoustic. The dispersion relation becomes 
\begin{equation}
\omega^2\simeq \epsilon_2\omega_{\rm p}^2+\epsilon_1\omega_{\rm p}^2\lambda_1^2k^2,
\end{equation}
so that the frequency is finite at $k=0$. 

\begin{figure}
\includegraphics[width=8.5cm]{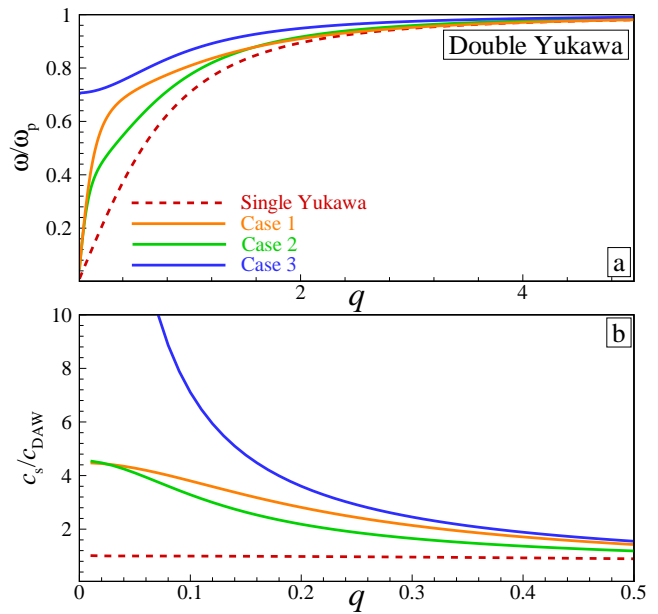}
\caption{(color online) The dispersion relation $\omega(q)$ (a) and the apparent sound velocity $c_{\rm s}(q)$ (b) of the double repulsive Yukawa potential (\ref{pot1}) in the weak coupling limit. Here the frequency is expressed in units of the plasma frequency scale $\omega_{\rm p}$ and the sound velocity is in units of the DAW sound velocity $c_{\rm DAW}=\omega_{\rm p}\lambda_{\rm D}$. The three solid curves correspond to the three potentials used in the calculations (Cases 1-3, see Table ~\ref{TabI} for details), as indicated in the figure. The dashed red curves correspond to the conventional DAW (single-Yukawa potential). }
\label{Fig2}
\end{figure}

The longitudinal mode dispersions for the double-Yukawa interaction potential in the weak coupling limit are shown in Fig.~\ref{Fig2}a. The three solid curves correspond to the three parameter sets considered (Cases 1, 2, and 3). The red dashed curve shows the corresponding dispersion for the single-Yukawa interaction potential. In Figure~\ref{Fig2}b we plot the apparent sound velocity $c_{\rm s}=\omega/k$, expressed in units of the conventional DAW sound velocity, $c_{\rm DAW}$ (the ''apparent'' in our context means that we retain the notion of sound speed, as defined above, even when the dispersion is non-acoustic). The important observation is that
the difference between the dispersion laws of the single-Yukawa and double-Yukawa potentials is most pronounced in the long-wavelength regime. The apparent acoustic velocity of the double-Yukawa system can exceed considerably the conventional DAW sound speed.         
      
\begin{figure}
\includegraphics[width=8.5cm]{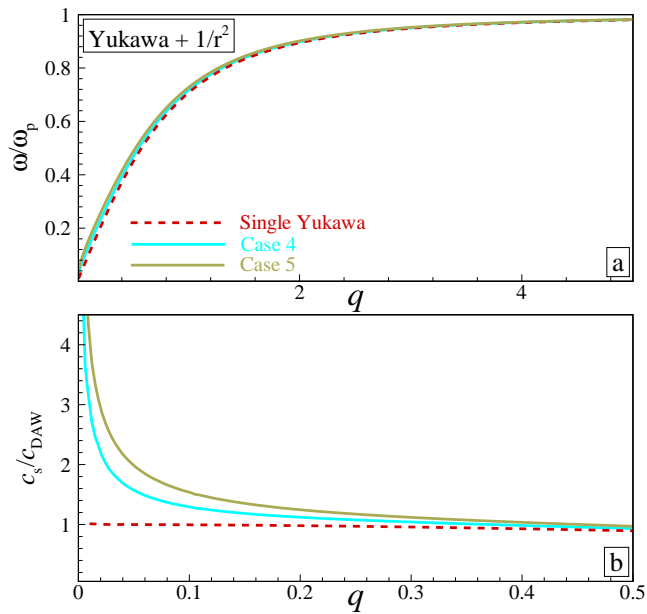}
\caption{(color online) The dispersion relation $\omega(q)$ (a) and the apparent sound velocity $c_{\rm s}(q)$ (b) of the repulsive Yukawa plus $1/r^2$ potential (\ref{pot2}) in the weak coupling limit. Here again the frequency is expressed in units of the plasma frequency scale $\omega_{\rm p}$ and the sound velocity in units of the DAW sound velocity $c_{\rm DAW}=\omega_{\rm p}\lambda_{\rm D}$. The two solid curves correspond to the Cases 4 and 5 (see Table~\ref{TabI} for details). The dashed red curves correspond to the conventional DAW (single-Yukawa potential).  }
\label{Fig3}
\end{figure}

For the potential (\ref{pot2}) in the weak coupling limit, the calculation yields (see Appenix for the details)
\begin{equation}\label{DispY+R}
\omega^2=\frac{(1-\epsilon)\omega_{\rm p}^2q^2}{q^2+\kappa^2}+\frac{\epsilon \omega_{\rm p}^2 q}{\kappa}\left[\frac{\pi}{2}-\arctan\left(\frac{q}{\kappa}\right)\right].
\end{equation}
Using the series expansions $\arctan(x)\simeq x + {\mathcal O}(x^3)$ for $x\rightarrow 0$ and $\arctan(x)\simeq \pi/2-1/x + {\mathcal O}(x^{-3})$ for $x\rightarrow \infty$ we get 
\begin{displaymath}
\omega\simeq \omega_{\rm p}
\end{displaymath} 
in the short-wavelength limit ($q\rightarrow\infty$) and   
\begin{displaymath}
\omega^2/\omega_{\rm p}^2\simeq \frac{\pi}{2}\epsilon k\lambda_{\rm D}+(1-2\epsilon)k^2\lambda_{\rm D}^2 
\end{displaymath} 
in the long-wavelength limit ($q\rightarrow 0$). The latter expression implies $\omega\propto \sqrt{k}$ at long wavelengths, i.e. non-acoustic character of the dispersion.   

The dispersion relations of the longitudinal mode for the weakly coupled system with the interaction potential~(\ref{pot2}) are shown in Fig.~\ref{Fig3}a. The solid curves correspond to the Cases 4 and 5, as indicated in the figure. The red dashed curve corresponds again to the single-Yukawa interaction potential. We observe that the dispersion relations themselves are not visually sensitive to the presence of the long-range unscreened $r^{-2}$ asymptote. However, the apparent acoustic velocity exceeds significantly the conventional DAW sound speed in the limit $q\ll 1$, as expected, since the apparent acoustic velocity diverges, $c_{\rm s}\propto k^{-1/2}$ as $k$ approaches zero.

\subsection{Strongly coupled regime} 
 
\begin{figure}
\includegraphics[width=8.5cm]{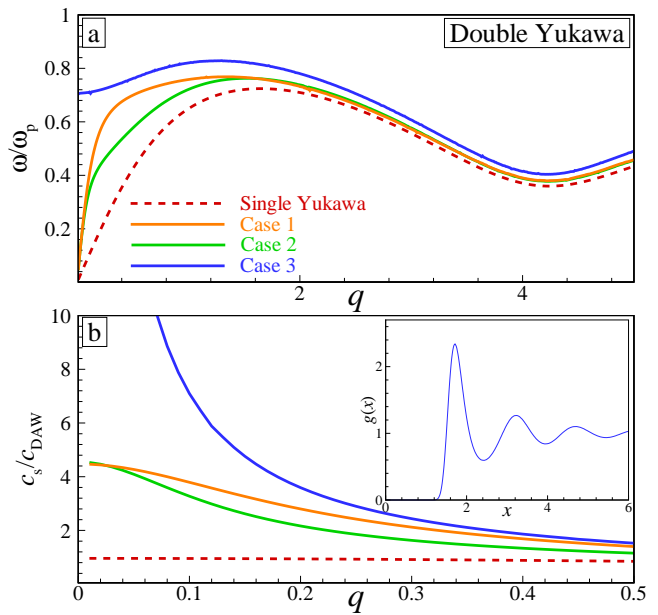}
\caption{(color online) The same as in Fig.~\ref{Fig2}, but in the strongly coupled regime (strong correlations between the particle positions). The inset in (b) shows the radial distribution function used to calculate the dispersion relations.}
\label{Fig4}
\end{figure}

As we pointed out in the introduction, the particle component in complex plasmas is often strongly coupled and forms condensed liquid and solid phases. Thus, dispersion relations derived above for the weakly coupled regime have limited applicability and should be supplemented by the respective relations for strongly coupled fluids. QCA model is a relevant tool for this purpose. In order to perform the calculation we have to use a realistic RDF $g(r)$ corresponding to the strongly coupled fluid regime. For the purpose of this study it is appropriate to take a single $g(r)$ for all the cases considered. This allows us to elucidate how the effect of strong coupling affects the properties of the dispersion relation in the most direct manner. The RDF employed here has been obtained using a standard molecular dynamics simulation for the particles interacting via the single-Yukawa potential and forming a strongly coupled fluid, very close to the fluid-solid phase transition~\cite{HamaguchiPRE1997}. The obtained RDF is plotted in the inset of Fig.~\ref{Fig4}b. We note in passing that in the regime of sufficiently strong coupling, the dispersion relations (in properly reduced units) are not very sensitive to the exact shape of the RDF and even simplistic models based on excluded volume arguments can provide reasonable results~\cite{KhrapakPoP2016}.

\begin{figure}
\includegraphics[width=8.5cm]{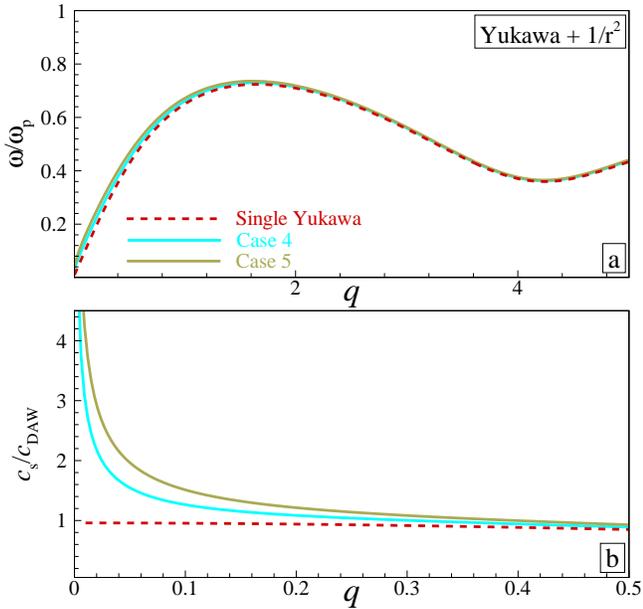}
\caption{(color online) The same as in Fig.~\ref{Fig3}, but in the strongly coupled regime (strong correlations between the particle positions). The RDF used in the calculations is the same as shown in the inset of Fig.~\ref{Fig4}b.}
\label{Fig5}
\end{figure}

Using the obtained $g(x)$ the dispersion curves of the longitudinal mode have been calculated with the help of Eq.~(\ref{Gen_expr}) from the appendix. The results for the double-Yukawa potential are presented in Fig.~\ref{Fig4}. Similar calculation for the Yukawa plus $r^{-2}$ long-range asymptote are depicted in Fig.~\ref{Fig5}. We observe the qualitative change of the dispersion curves compared to the weakly coupled regime. The frequency does not increase monotonically to reach the asymptotic value of $\omega_{\rm p}$ in the short-wavelength limit. Instead, the frequency reaches a maximum (at $q\lesssim 2$) whose magnitude is below $\omega_{\rm p}$. At larger $q$ the frequency is known to exhibit a series of damped oscillations on approaching the short-wavelength asymptote -- the Einstein frequency~\cite{KalmanPRL2000}. On the other hand, we see from Figs.~\ref{Fig4} and \ref{Fig5} that the behaviour of the apparent acoustic velocity has not changed much compared to the weakly coupled regime. This is merely a consequence of the condition $\kappa=1$ used in our calculations. It has been reported that the ratio $c_{\rm s}/c_{\rm DAW}$ in strongly coupled Yukawa systems is rather close to unity at $\kappa\lesssim 1$, but then drops considerably as $\kappa$ increases further (for instance, $c_{\rm s}/c_{\rm DAW}\sim 0.3$ at $\kappa=5$)~\cite{KhrapakPRE03_2015,KhrapakPPCF2016}. Thus, some quantitative differences between the sound speeds in weakly and strongly coupled regimes should be expected upon an increase in $\kappa$. However, this will not affect the main point of our present study -- qualitative and quantitative differences in the waves dispersion arising due to deviation from the single-Yukawa interaction potential. In particular, it is observed that the apparent sound speed can increase considerably compared to the conventional DAW value when repulsive long-range modifications to the single-Yukawa potential are present. In addition, $c_{\rm s}$ exhibits significant negative slope in the low-$q$ domain, while for the single-Yukawa potential it remains practically constant.

\subsection{Effect of neutral gas damping}

The QCA (QLCA) theory excludes consideration of various damping effects. One damping effect, particularly relevant for complex plasmas is associated with the collisions between charged dust particles and neutral atoms or molecules (ion-particle and electron-particle collisions also take place, but in typical weakly ionized gas discharges neutral damping dominates). Although the damping is relatively weak under typical experimental conditions it is inevitably present in experiments. An important question is, therefore, to which extent it can affect the results derived so far.     

\begin{figure}
\includegraphics[width=8.5cm]{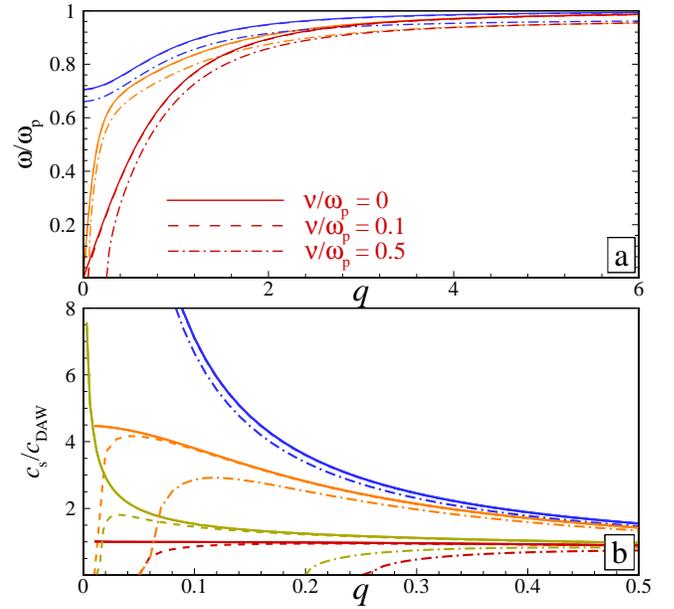}
\caption{(color online) Effect of the neutral gas damping on the dispersion relation of the longitudinal waves (a) and the apparent sound velocity (b) in the weakly coupled regime. The dispersion relations are shown for the cases 1 (orange),  3 (blue) and the single-Yukawa potential (red). The sound velocities are shown for the cases 1, 3, 5 (olive) and the single-Yukawa potential. The solid, dotted, and dash-dotted curves correspond to the damping rates $\nu/\omega_{\rm p} =0$, $\nu/\omega_{\rm p}=0.1$, and $\nu/\omega_{\rm p}=0.5$, respectively.}
\label{Fig6}
\end{figure}

The effect of damping can be included in an {\it ad hoc} manner and results in the replacement $\omega^2\rightarrow \omega (\omega+i\nu)$ in Eq.~(\ref{QCA})~\cite{RosenbergPRE1997,HouPRE2009,RosenbergPRE2014}, where $\nu$ is the  damping rate due to collisions with neutrals. The magnitude of the damping rate can be varied considerably, in particular adjusting the neutral gas pressure. For the neutral gas pressures in the range between $\sim 10$ Pa and $\sim 50$ Pa the reduced collisional damping rates were estimated in the range $\nu/\omega_{\rm p}\simeq 0.2-0.3$ in different experiments with low-frequency dust waves described in Refs.~\cite{KhrapakPoP2003,RatynskaiaIEEE2004,YaroshenkoPRE2004,PielPRE2008}. In Ref.~\cite{BandyopadhyayPLA2007} the reduced damping rate varied between  $\nu/\omega_{\rm p}\simeq 0.07$ at a pressure $p=8.6$ Pa and  $\nu/\omega_{\rm p}\simeq 0.6$ at $p=50$ Pa. In general, in addition to pressure, the reduced damping rate depends on a number of system parameters (e.g.,  particle size, charge, mass, and number density, gas type, etc.) However, the values listed above can be considered as representative.             
Here we take two values, $\nu/\omega_{\rm p}\simeq 0.1$ (weak damping) and $\nu/\omega_{\rm p}\simeq 0.5$ (strong damping) and recalculate the dispersion relations derived above taking into account the damping effect.

\begin{figure}
\includegraphics[width=8.5cm]{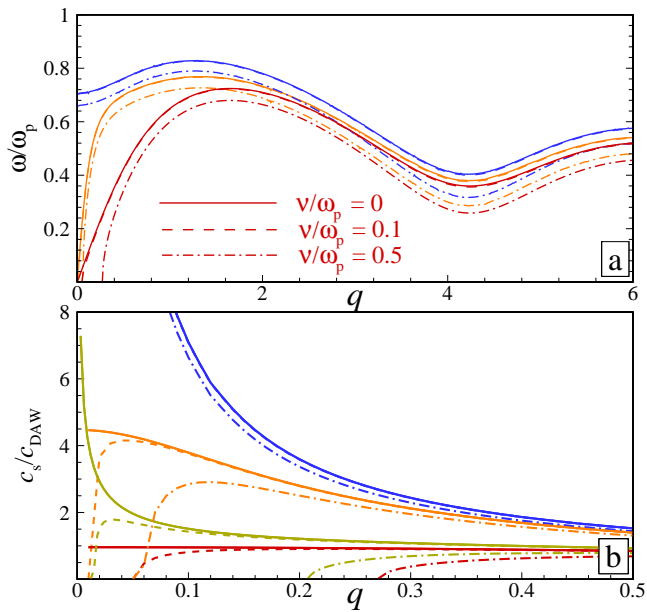}
\caption{(color online) Same as in Figure \ref{Fig6}, but for the strongly coupled regime.}
\label{Fig7}
\end{figure}

The results are presented in Figs.~\ref{Fig6} and \ref{Fig7}. The first of them corresponds to the weakly coupling regime, while the second to the strongly coupled regime. The dispersion relations are hardly affected by weak damping. One cannot see the difference between the curves corresponding to $\nu/\omega_{\rm p}=0$ and $\nu/\omega_{\rm p}=0.1$ on the scale of Figs.~\ref{Fig6}a and \ref{Fig7}a. At $\nu/\omega_{\rm p}=0.5$ the difference becomes more pronounced: The frequencies are somewhat shifted down. Collisional effects are expected to dominate at long-wavelengths since in this regime the wave frequencies can be low. Therefore, in Figs.~\ref{Fig6}b and \ref{Fig7}b we show the apparent sound velocities in the long-wavelength regime. It is seen that collisions can have considerable effect on the wave propagation for the cases when $\omega\rightarrow 0$ at $k\rightarrow 0$. In contrast, when the frequency starts from a finite value at $k=0$ (Case 3) the collisional effects are seen insignificant. Overall, we can summarize this Section as follows. Neutral damping affects mostly the long-wavelength part of the dispersion. This is exactly where the deviations from the single-Yukawa potential of interaction can dominate the dispersion relation. To single out the latter effect in the experiments one therefore needs to reduce collisional effects (e.g. by lowering the neutral gas pressure and/or adjusting other complex plasma parameters, see Ref.~\cite{KalmanPRL2000} for a relevant discussion).

\section{Discussion and Conclusion}\label{Disc}

One of the most important conclusions from this study is that the conventional dispersion relation of the dust acoustic waves (DAW) is not an inherent property of complex (dusty) plasmas. The DAW dispersion operates when the interparticle interactions are of Yukawa (screened Coulomb) form. Deviations from the Yukawa form result in deviations in the dispersion law. 

In order to demonstrate this we have used the quasi-crystalline approximation and derived the corresponding    
dispersion relations for the longitudinal waves for several representative pair interaction potentials, which can be operable in complex plasmas. The interaction considered include double-Yukawa, Yukawa plus long-range Coulomb asymptote, and Yukawa plus long-range $r^{-2}$ asymptote (all repulsive). Both, weakly coupled and strongly coupled regimes have been studied.

The obtained results demonstrate how the variations in the interparticle interaction potential affect the dispersion relation. In particular, the long-range asymptotic behaviour of the potential determines the long-wavelength behaviour of the dispersion relation. A useful measure of the deviations is the apparent sound velocity, $c_{\rm s}=\omega/k$. This quantity remains practically constant for the single-Yukawa potential, at least in the regime $q=ka\lesssim 0.5$, and is given by the DAW sound velocity, $c_{\rm DAW}$, at weak coupling. In the strongly coupled regime, it is also close to $c_{\rm DAW}$ when screening is weak ($\kappa\lesssim 1$), but decreases when screening strengthens. When repulsive long-range asymptotes are present, the apparent sound velocity can increase considerably, compared to the single-Yukawa case, and demonstrates significant negative slope in the same range of $q$. This can be in principle used to verify the existence of deviations from the conventional Yukawa interactions in complex plasmas experimentally.          

Experimental observations of dust acoustic waves have a long-standing history~\cite{BarkanPoP1995,PieperPRL1996,ThompsonPoP1997,MerlinoPoP1998,FortovPoP2000,
RatynskaiaPRL2004,PielPRL2006,ThomasJrPoP2007,MerlinoPPCF2012}. Most of the available observations correspond to the long-wavelength regime, $q\lesssim 1$. To the best of our knowledge, however, the experimental results were not analysed from the point of view of inferring that interactions in complex plasmas can deviate from the conventional single-Yukawa form. The theoretical results presented here can be useful in this context as they provide guidelines for new dedicated experiments. The two most important requirements for such experiments identified here are the accurate resolution of the longitudinal dispersion relation in the long-wavelengths limit and sufficiently weak collisionality. In this case, careful analysis should be able to discriminate between different long-range asymptotes predicted theoretically, or at least validate their existence.

\begin{acknowledgments}
This work was supported by the A*MIDEX project (Nr.~ANR-11-IDEX-0001-02) funded by the French Government ``Investissements d'Avenir'' program managed by the French National Research Agency (ANR). It was also partially supported by the French-German PHC PROCOPE Program (Project No. 35325NA/57211784). Simulations were supported by the Russian Science Foundation (grant Nr. 14-12-01185).
\end{acknowledgments}

\appendix

\section{Dispersion relations at weak coupling}

We assume that a pairwise interaction potential can be written in the form
\begin{displaymath}
V(r)=\varepsilon f(r/a),
\end{displaymath} 
where $\varepsilon$ is the energy scale. Then, the generic (QCA) expression for the longitudinal wave dispersion relation in 3D resulting from (\ref{QCA}) is
\begin{widetext}
\begin{equation}\label{Gen_expr}
\omega^2= \omega_0^2\int_0^{\infty} xg(x)dx\left\{f'(x)\left[\frac{2}{3}+\frac{2\cos qx}{q^2 x^2}-\frac{2\sin qx}{q^3x^3}\right]+xf''(x)\left[\frac{1}{3}+\frac{2\sin qx}{q^3x^3}-\frac{2\cos qx}{q^2 x^2}-\frac{\sin qx}{qx}\right]
\right\},
\end{equation}
\end{widetext}
where $\omega_0^2=4\pi n\varepsilon a/m$ is the nominal frequency. For the potentials considered here $\varepsilon=Q^2/a$ and the nominal frequency coincides with the conventional plasma frequency, $\omega_0=\omega_{\rm p}$. In the weakly coupled limit the correlations between the particles positions are absent and we can put $g(x)=1$ into Eq.~(\ref{Gen_expr}), which corresponds to the random phase approximation~\cite{Hubbard1969}. Note that in order the integral in Eq.~(\ref{Gen_expr}) converges at small $x$, the potential should generally rise slower than $\propto x^{-3}$ when $x\rightarrow 0$, which is the case for the potentials studied here.         

For the single Yukawa potential we have $f(x)=e^{-\kappa x}/x$ and the integration can be done analytically. The result corresponds to the conventional DAW dispersion relation of Eq.~(\ref{DAW}). 

Next, consider the potential of the form $f(x)=e^{-\kappa x}/x^2$. The integration can again be done analytically and yields
\begin{equation}
\omega^2=\omega_0^2q \arctan(q/\kappa).
\end{equation} 
In the unscreened limit ($\kappa=0$) we get
\begin{equation}
\omega^2=\frac{1}{2}\pi q \omega_0^2,
\end{equation}
which is the dispersion relation for the $f(x)=1/x^2$ interaction in the limit of weak coupling. Using these results, Eq.~(\ref{DispY+R}) is readily obtained.

\bibliographystyle{aipnum4-1}
\bibliography{KhrapakPRE_Aug2016}

\end{document}